\documentclass[12pt]{article}

\pdfoutput=1
\usepackage{graphicx} 
\usepackage{xspace}
\usepackage{url}
\usepackage{comment}
\usepackage{amsmath}
\usepackage{amssymb}
\usepackage{fancyhdr}  
\usepackage{hepunits}
\usepackage[usenames,dvipsnames]{color}  
\usepackage{enumitem}
\usepackage{multirow}
\usepackage{lineno}
\usepackage{pdfpages}
\usepackage{wasysym}
\usepackage{parskip}
\usepackage[a4paper, total={17cm, 24cm}, headheight=0.6cm]{geometry}
\usepackage{subcaption}  
\usepackage[pdftex,bookmarks,hidelinks]{hyperref}

\usepackage[normalem]{ulem}
\def\phaseone{Phase~I\xspace} 
\def\phasetwo{Phase~II\xspace} 

\newcommand{\rms}{RMS\xspace} 

\newcommand{\deltacp}{\ensuremath{\delta_{\rm CP}}\xspace}   

\def\argon40{${}^{40}$Ar}       
\def\Ar39{$^{39}$Ar}
\def\Cl40{$^{40}$Cl}
\def\K40{$^{40}$K}
\def\B8{$^{8}$B}
\newcommand{\lsim}{{\;\raise0.3ex\hbox{$<$\kern-0.75em\raise-1.1ex\hbox{$\sim$}}\;}}
\newcommand{\gsim}{{\;\raise0.3ex\hbox{$>$\kern-0.75em\raise-1.1ex\hbox{$\sim$}}\;}}
\newcommand{\beq}{\begin{equation}}
\newcommand{\eeq}{\end{equation}}
\newcommand{\bea}{\begin{eqnarray}}
\newcommand{\eea}{\end{eqnarray}}

\mathchardef\minus="002D

\DeclareSIUnit \c {$c$}
\DeclareSIUnit\magn{$\times$}
\DeclareSIUnit\min{min}
\DeclareSIUnit\hr{hr}
\DeclareSIUnit\hrs{hrs}
\DeclareSIUnit\week{week}
\DeclareSIUnit\month{mo}
\DeclareSIUnit\months{mos}
\DeclareSIUnit\year{yr}
\DeclareSIUnit\years{years}
\DeclareSIUnit\yr{yr}
\DeclareSIUnit\standard{std}
\DeclareSIUnit\str{sr}
\DeclareSIUnit\ppm{ppm}
\DeclareSIUnit\ppb{ppb}
\DeclareSIUnit\ppt{ppt}
\DeclareSIUnit\pe{PE}
\DeclareSIUnit\spe{SPE}
\DeclareSIUnit\pdm{PDM}
\DeclareSIUnit\ev{events}
\DeclareSIUnit\ct{counts}
\DeclareSIUnit\neutron{\mbox{$n$}}
\DeclareSIUnit\smp{samples}
\DeclareSIUnit\Sample{S}
\DeclareSIUnit\ch{ch}
\DeclareSIUnit\hit{hit}
\DeclareSIUnit\hits{hits}
\DeclareSIUnit\bin{(\mbox{5-PE}~bin)}
\DeclareSIUnit\sgm{\mbox{$\sigma$}}
\DeclareSIUnit\rms{RMS}
\DeclareSIUnit\keVee{\mbox{keV$_{e{\rm e}}$}}
\DeclareSIUnit\keVr{\mbox{keV$_{\rm nr}$}}
\DeclareSIUnit\eVee{\mbox{eV$_{\rm ee}$}}
\DeclareSIUnit\eVr{\mbox{eV$_{\rm nr}$}}
\DeclareSIUnit\ph{photon}
\DeclareSIUnit\el{\mbox{$e^-$}}
\DeclareSIUnit\pm{\mbox{PMT}}
\DeclareSIUnit\pixel{\mbox{pixel}}
\DeclareSIUnit\inch{''}
\DeclareSIUnit\foot{'}
\DeclareSIUnit\bit{bit}
\DeclareSIUnit\sample{samples}
\DeclareSIUnit\barn{barn}
\DeclareSIUnit\bara{bar}
\DeclareSIUnit\bar{bar}
\DeclareSIUnit\barg{barg}
\DeclareSIUnit\mlardepth{\mbox(meter~of~\LAr~depth)}
\DeclareSIUnit\Curie{Ci}
\DeclareSIUnit\PSI{psi}
\DeclareSIUnit\psia{psia}
\DeclareSIUnit\atm{atm}
\DeclareSIUnit\psf{psf}
\DeclareSIUnit\pcf{pcf}
\DeclareSIUnit\parsec{pc}
\DeclareSIUnit\cps{cps}
\DeclareSIUnit\slpm{\SI{}{\liter\per\minute}}
\DeclareSIUnit\rpm{rpm}
\DeclareSIUnit\mwe{\mbox{m.w.e.}}
\DeclareSIUnit\liveday{\mbox{live-days}}
\DeclareSIUnit\days{\mbox{days}}
\DeclareSIUnit\miles{\mbox{miles}}
\DeclareSIUnit\lumens{\mbox{lm}}
\DeclareSIUnit\degreeC{\mbox{$^{\circ}$C}}
\DeclareSIUnit\degreeF{\mbox{$^{\circ}$F}}
\DeclareSIUnit\electron{\mbox{$e^-$}}
\DeclareSIUnit\Euro{\mbox{\euro}}
\DeclareSIUnit\cph{cph}
\DeclareSIUnit\neq{neq}
\DeclareSIUnit\normal{\mbox{N}}
\DeclareSIUnit\USD{\mbox{\$}}
\DeclareSIUnit\Vpercm{\mbox{V/cm}}
\DeclareSIUnit\kV{\mbox{kV}}

\DeclareSIUnit \mm {\milli\meter}
\DeclareSIUnit \cm {\centi\meter}
\DeclareSIUnit \us {\micro\second}
\DeclareSIUnit \ms {\milli\second}
\DeclareSIUnit \pA {\pico\ampere}
\DeclareSIUnit \pC {\pico\coulomb}
\DeclareSIUnit \fC {\femto\coulomb}
\DeclareSIUnit \fF {\femto\farrad}
\DeclareSIUnit \pF {\pico\farrad}
\DeclareSIUnit \mV {\milli\volt}
\DeclareSIUnit \kV {\kilo\volt}
\DeclareSIUnit \V {\volt}
\DeclareSIUnit \GOhm {\giga\ohm}
\DeclareSIUnit \MOhm {\mega\ohm}
\DeclareSIUnit \ton {\tonne}
\DeclareSIUnit \kton {\kilo\tonne}
\DeclareSIUnit \kt {\kilo\tonne}
\DeclareSIUnit \Mt {\mega\tonne}
\DeclareSIUnit \eV {\electronvolt}
\DeclareSIUnit \keV {\kilo\electronvolt}
\DeclareSIUnit \MeV {\mega\electronvolt}
\DeclareSIUnit \GeV {\giga\electronvolt}
\DeclareSIUnit \km {\kilo\meter}
\DeclareSIUnit \kW {\kilo\watt}
\DeclareSIUnit \MW {\mega\watt}
\DeclareSIUnit \MHz {\mega\hertz}
\DeclareSIUnit \kHz {\kilo\hertz}
\DeclareSIUnit \mrad {\milli\radian}
\DeclareSIUnit \year {year}
\DeclareSIUnit \POT {POT}
\DeclareSIUnit \sig {$\sigma$}
\DeclareSIUnit\parsec{pc}
\DeclareSIUnit\lightyear{ly}
\DeclareSIUnit\foot{ft}
\DeclareSIUnit\ft{ft}

\title{%
European Contributions to Fermilab Accelerator Upgrades and Facilities for the DUNE Experiment\\ \bigskip
\large Input to the European Strategy for Particle Physics - 2026 Update }
\author{The DUNE Collaboration
\footnote{Contact persons: Sergio Bertolucci (Sergio.Bertolucci@cern.ch), Sowjanya Gollapinni (sowjanya@lanl.gov)}
}
\date{\today}

\begin{document}

\maketitle

\begin{abstract}
    The Proton Improvement Plan (PIP-II) to the FNAL accelerator chain and the Long-Baseline Neutrino Facility (LBNF) will provide the world's most intense neutrino beam to the Deep Underground Neutrino Experiment (DUNE) enabling a wide-ranging physics program. This document outlines the significant contributions made by European national laboratories and institutes towards realizing the first phase of the project with a 1.2 MW neutrino beam. Construction of this first phase is well underway. For DUNE \phasetwo, this will be closely followed by an upgrade of the beam power to $>2$~MW, for which the European groups again have a key role and which will require the continued support of the European community for machine aspects of neutrino physics. 
    
    Beyond the neutrino beam aspects, LBNF is also responsible for providing unique infrastructure to install and operate the DUNE neutrino detectors at FNAL and at the Sanford Underground Research Facility (SURF). The cryostats for the first two Liquid Argon Time Projection Chamber detector modules at SURF, a contribution of CERN to LBNF, are central to the success of the ongoing execution of DUNE \phaseone. Likewise, successful and timely procurement of cryostats for two additional detector modules at SURF will be critical to the success of DUNE \phasetwo and the overall physics program. 

    The DUNE Collaboration is submitting four main contributions to the 2026 Update of the European Strategy for Particle Physics process. This paper is being submitted to the “Accelerator technologies” and “Projects and Large Experiments” streams. Additional inputs related to the DUNE science program, DUNE detector technologies and R\&D, and DUNE software and computing, are also being submitted to other streams. 
\end{abstract}

\thispagestyle{empty} 

\newpage
\pagenumbering{arabic}

\section{Introduction/Context}
\label{sec:context}
\addcontentsline{toc}{section}{Executive summary}

Understanding the origin of the asymmetry of matter over antimatter in our universe and the nature of the new physics responsible for neutrino mass are priorities within the particle physics community. This has been recognized by the conclusions of the 2013 update of the European Strategy for Particle Physics~\cite{2013europeanstrategy}, the 2014 Report of the US Particle Physics Project Prioritization Panel (P5)~\cite{2014p5report} and the subsequent update of the P5 in 2023~\cite{2023p5report}. The Deep Underground Neutrino Experiment (DUNE) was launched in 2015 to address these questions with a long baseline, broadband, neutrino oscillation experiment. With state-of-the-art liquid argon technology, DUNE is also capable of a much broader physics program including nucleon decay searches, as well as neutrino physics and astrophysics measurements using solar, galactic supernova and atmospheric neutrino probes.      

DUNE is an international collaboration tasked with designing and constructing a near detector complex of detectors based at Fermi National Accelerator Laboratory (FNAL) and a suite of four far detectors located at the Sanford Underground Research Facility (SURF) 1300~km north-west of FNAL. The neutrino beamline, together with all excavations and infrastructure to house and support the DUNE detectors, is provided by the Long Baseline Neutrino Facility (LBNF). 

LBNF-DUNE will be implemented in two phases. The first phase will consist of a near detector (ND), two far detector (FD) modules (giving $>20$~kt of fiducial mass) and a proton beam power of $1.2$~MW. Construction of this phase is now well under way and will begin first beam physics in 2031. \phasetwo of the project will roll out two additional far detectors (giving in total $>40$~kt LAr equivalent fiducial mass), an upgraded ND and beam power over $2$~MW, to be complete early in the next decade as fast as resources allow.   

The LBNF-DUNE project in the USA is responsible for providing much of the facility and infrastructure needed at FNAL and SURF. There are, however, many key contributions from international partners without which the project would not be viable -- the DUNE collaboration  has $53$\%~non-US collaborators, 39$\%$~of whom come from European institutes. In addition to delivering detectors and related infrastructure, Europe is making key contributions to LBNF and the FNAL accelerator upgrade. This document gives a brief overview and status of European contributions to the FNAL accelerator upgrade and to detector infrastructures at SURF. The success of LBNF-DUNE relies heavily on the European community continuing to support  aspects of the facility. Furthermore, developing capability in these areas, for example in the design of superconducting radio frequency (SRF) cavities, is going to be a key asset underpinning planning in Europe for future linear or circular colliders at CERN.

\section{Accelerator Upgrades Program}
\label{sec:ace}
\begin{figure}[!tb]
  \centering
  \raisebox{0.5cm}{\includegraphics[width=0.55\textwidth]{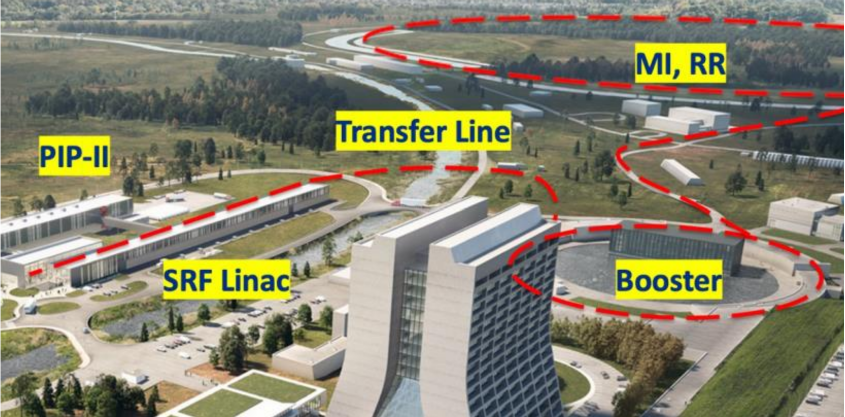}}
  \includegraphics[width=0.4\textwidth]{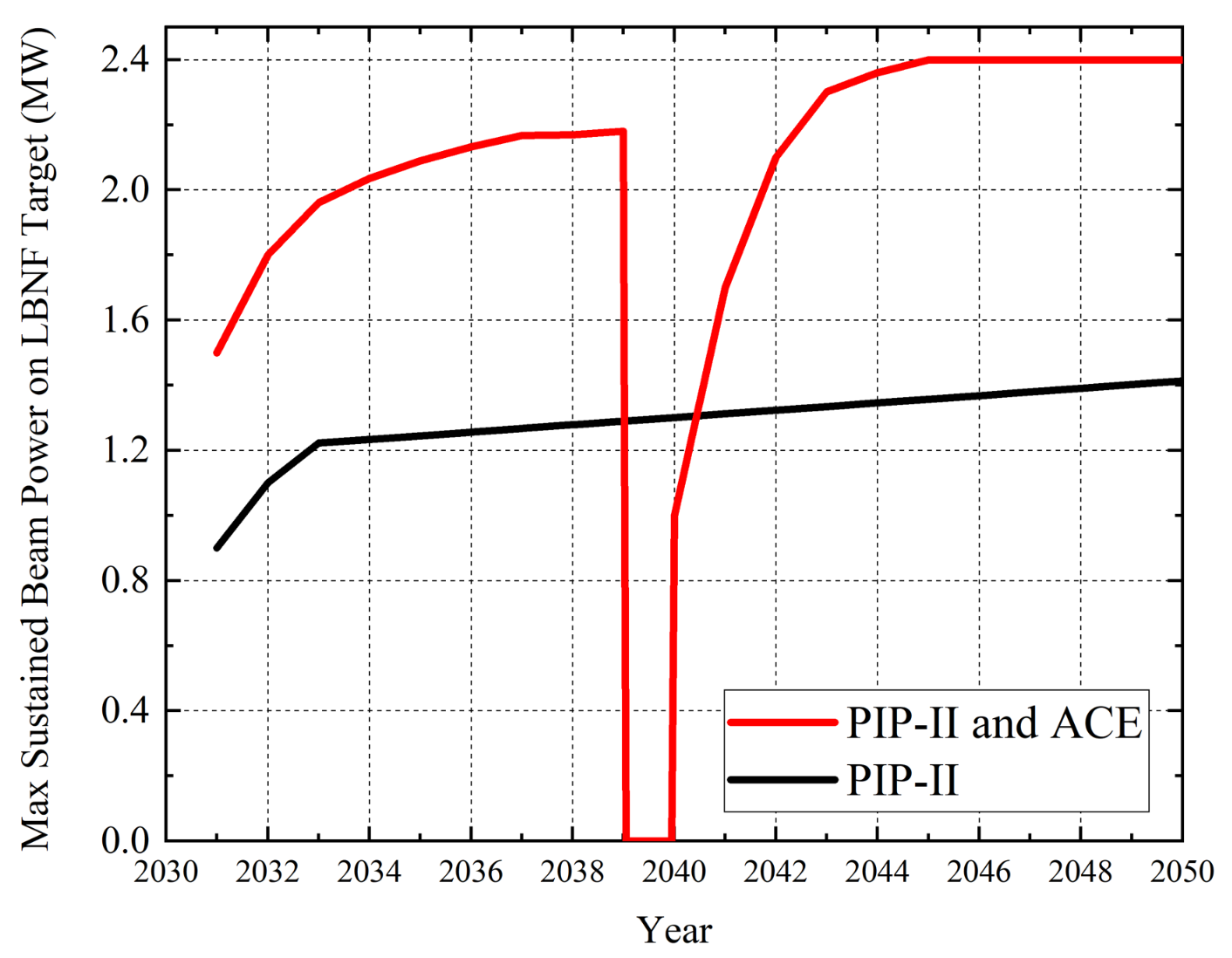}
  \caption{\small 
    (Left) Elements of the PIP-II accelerator upgrade at FNAL. (Right) Maximum sustained beam power on the LBNF neutrino target under two scenarios: no upgrade beyond PIP-II is shown by the black line, while PIP-II with ACE-MIRT (2031--2039) and ACE-BR (starting in 2040) upgrades is shown in red.}
  \label{FIG-ACE}
\end{figure}

In order to deliver the proton beam requirements for DUNE and to satisfy the future needs of the muon program and other applications, FNAL is embarking on an extensive upgrade of its accelerator complex. The first stage of this is the Proton Improvement Plan (PIP-II~\cite{PIP2Concept}) which is providing a new SRF linear accelerator for injection into the Booster at $800$~MeV, a new Beam Transfer Line from the Linac to the Booster and the upgrade of components in the Booster, Recycler Ring (RR) and Main Injector (MI), see Fig.~\ref{FIG-ACE} (Left). These changes will enable a proton beam of $1.2$~MW for the start of beam physics with DUNE \phaseone.

An increase in beam power to $> 2$~MW is a primary requirement for DUNE to achieve its P5 physics goals. Fermilab has in place an Accelerator Complex Evolution (ACE) plan to realize the beam power required. It will be delivered in two phases. The first phase, known as ACE-Main Injector Ramp and Targetry (ACE-MIRT), will raise the beam power to just beyond $2$~MW by reducing the cycling time in the MI from $1.2$~sec to $0.7$~sec. This is to be followed by the replacement of the Booster and an extension of the Linac to $2$~GeV (ACE-BR), which will eventually allow the beam power to rise to $2.4$~MW. The importance of these upgrades for faster and more reliable delivery of protons on target can be seen in Fig.~\ref{FIG-ACE} (Right)~\cite{2672528}. The integrated DUNE exposure over the 2031--2050 period is expected to increase by approximately 50\% thanks to the ACE upgrades alone. While the ACE-MIRT upgrade is primarily motivated by DUNE science, ACE-BR targets a broader physics program at FNAL, serving as a robust and reliable platform for the future.

An activity, in the US and the UK, has begun to develop a high power target system capable of operating at a beam power in excess of $2$~MW. It demands an increase in the cooling capacity of the target design for $1.2$~MW and materials studies to investigate suitable target and component infrastructure materials. Not all resources are currently in place to make the upgrades needed to the current target design. It is going to be crucial for UK support to be continued past the existing PIP-II target project (ending in 2028).

\section{European Contributions to PIP-II}
\label{sec:PIP-II}
\begin{figure}[!tb]
  \centering
  \includegraphics[trim={0 0 0 0},clip,width=0.6\textwidth]{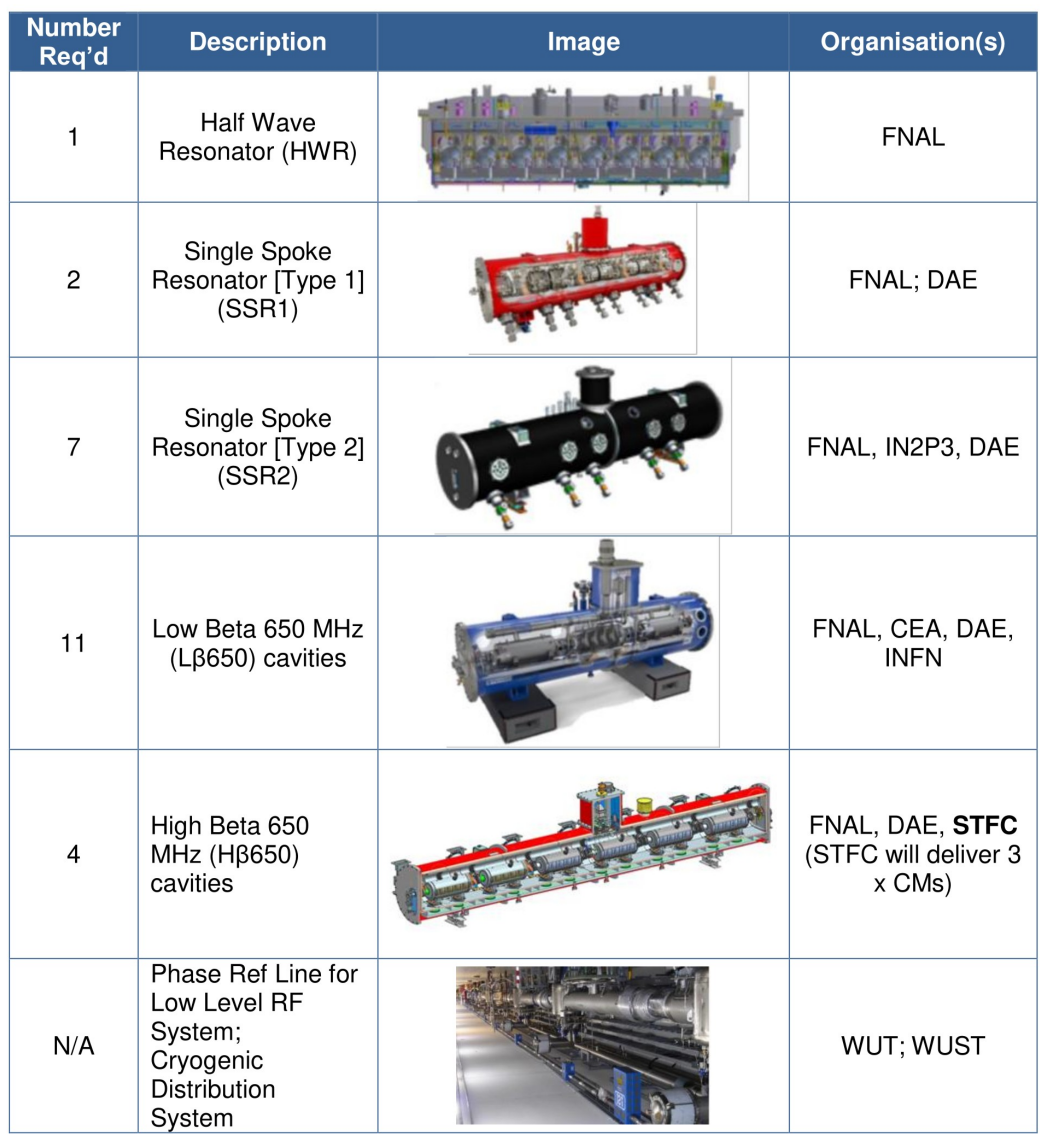}
  \caption{\small 
    International partner contributions to PIP-II: USA, Fermi National Accelerator Lab (FNAL); France, Commissariat  \`a l\'energie atomique et aux \'energies alternatives (CEA); India, Department of Atomic Energy (DAE); Italy, Istituto Nazionale di Fisica Nucleare (INFN); Poland, Wroclaw University of Science and Technology (WUST); UK, Science and Technology Facilities Council (STFC).
  }
  \label{FIG-PIP2Contribs}
\end{figure}

The 2014 P5 recommended that the Fermilab proton accelerator complex be upgraded via PIP-II to provide proton beams of \>1.2 MW ready for first operation of the LBNF. The development and construction of the project is relying on the considerable expertise in SRF technology that exists in the European community:
\begin{itemize}
    \item UK, UKRI: Substantial engineering, manufacturing and qualification experience through the construction and operation of light and neutron sources plus SRF cavity processing and testing for the European Spallation Source (ESS).
    \item Italy, INFN: Internationally recognized leader in SRF technologies, having provided SRF cavity and cryomodule fabrication for the European X-Ray Free-Electron Laser Facility (XFEL) and SRF cavities for the ESS.
    \item France, CEA, CNRS/IN2P3: Internationally recognized leader in large-scale cryomodule assembly through involvement with XFEL and the Single Spoke Resonators type 2 (SSR2) cavities and couplers for ESS. 
    \item Poland, WUST: Substantial engineering and manufacturing experience in cryogenic distribution and low level radio frequency systems, with contributions to XFEL and ESS.  
\end{itemize}
The PIP-II proton beam is provided by Radio Frequency (RF) accelerators which incrementally increase the beam energy through a series of normal conducting and SRF technology linear accelerator stages. The centerpiece is a new $800$ MeV superconducting proton linac (replacing a $400$ MeV
normal-conducting machine). The accelerator chain consists of a normal conducting ion source, RF Quadrupole (RFQ) and Medium Energy Beam Transport (MEBT) systems, superconducting Half Wave Resonator (HWR) and Spoke (SSR1 and SSR2) linac sections culminating with superconducting elliptical low beta (LB650) and high beta (HB650) cryomodules.     
Fig.~\ref{FIG-PIP2Contribs} gives an overview of the partner responsibilities to deliver the various components of the linac. 

PIP-II is being run essentially as a DOE project, with international contributions through partnerships. 
There are major European contributions which have enabled the project schedule to advance by reducing the funding required from the US DOE and also mitigated competition for resources within FNAL. 

Production runs to deliver the components of PIP-II described in Fig.~\ref{FIG-PIP2Contribs} are now well underway.
\begin{itemize}
\item {\bf SSR2 Single Spoke Resonator}: The IN2P3 team at IJCLAB have responsibility for delivering 33 accelerating cavities for the SSR2 modules, designed jointly with FNAL and procured from the Italian company, Zanon Research and Innovation SRL - who are also the supplier of cavities for the LB650 and HB650 modules. Challenging performance specifications have demanded the development of careful surface treatment techniques and, to date, the project has validated 5 cavities with performance well above specification. This work is due to complete in 2026.   
\item {\bf LB650 Low Beta Cryomodules}: CEA Saclay are responsible for the LB650 cryomodule integration, utilizing qualified SRF cavities provided by INFN-Milan. The LB650 cryomodules each contain four LB650 cavities (procured from Zanon) with an integrated cryomodule design that is identical to the HB650 cryomodule, leading to close collaboration between the STFC, Milan and CEA groups. CEA are preparing to start cryomodule assembly in the next 12 months.  
\item {\bf HB650 High Beta Cryomodules}: Four cryomodules are required, each consisting of six accelerating cavities. At UKRI's STFC-Daresbury Lab, its ASTeC and Technology department will deliver three HB650 cryomodules with the remaining one assembled at FNAL. Assembly of the first module in the UK will start in 2025 within the Superconducting RF Laboratory (SuRFLab), where new clean room facilities to assemble cavity strings and integrate into cryomodule vessels are close to completion. Promising results from pre-series Zanon cavities have been obtained, including a new world record performance for cavities of this type. The delivery of the first cryomodule to FNAL is on-course to be mid 2026 and all three modules delivered by mid-2027.   
\item {\bf Cryogenic Distribution System}: The PIP-II linac Cryogenic Distribution System (CDS) is tasked with providing cooling power from the cryogenic plant to 23 cryomodules. CDS consists of a Distribution Valve Box (DVB), Intermediate Transfer Line, Tunnel Transfer Line, comprising 25 Bayonet Cans, and ends with a Turnaround Can. CDS is characterized by extremely small heat inflows and robust mechanical design. The CDS geometry allows maintaining an acceptable pressure drop for each helium stream, considering the planned flows and helium parameters in different operation modes. This is particularly crucial for the return line of helium vapors, which return from cryomodules to the cold compressors and thus have very restrictive pressure drop requirements. 
\end{itemize}

\section{European Contributions to LBNF}
\label{sec:LBNF}
\begin{figure}[!tb]
  \centering
  \includegraphics[width=\textwidth]{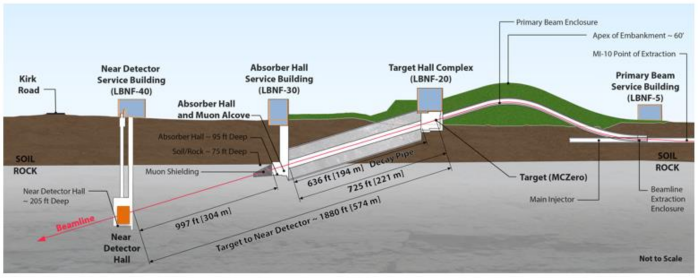}
  \caption{\small 
    The LBNF infrastructure at FNAL.
  }
  \label{FIG-LBNF}
  \end{figure}
The Long Baseline Neutrino Facility (LBNF) consists of a proton beamline capable of delivering $1.2$ MW of beam power from the Main Injector over an energy range of $60-120$ GeV, a Neutrino Target Station and associated infrastructure, plus the underground facilities necessary to host the DUNE Near and Far Detector systems. Figure~\ref{FIG-LBNF} shows the LBNF infrastructure at the FNAL site, which directs a neutrino beam towards the Far Detector site located at the SURF lab in South Dakota. 

Current contributions to LBNF from the European funding bodies are:
\begin{itemize}
\item CERN: Cryostats for the first two far detector modules of DUNE, plus associated  cryogenic infrastructure (cryo-mezzanine supports and surface liquid argon receiving tanks). 
\item UKRI/STFC/RAL: Design and delivery of a high power neutrino target for LBNF plus associated infrastructure.
\item Switzerland: Argon condenser system. 
\end{itemize}

\begin{figure}[!tb]
  \centering
  \includegraphics[width=\textwidth]{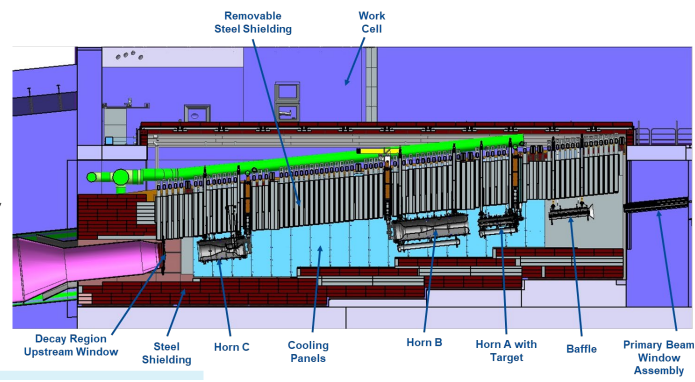}
  \caption{\small 
    The LBNF Neutrino Target Station.
  }
  \label{FIG-Beamline}
\end{figure}

Most European R\&D contributions to LBNF have centered on the LBNF Neutrino Beamline. This is a set of components designed to efficiently convert an intense proton beam into a high intensity neutrino beam aimed at the DUNE far detectors. Key design considerations include the need to provide a wideband neutrino beam to cover the first and second neutrino oscillation maxima and a path to upgrade the primary beam power from $1.2$ MW to $2.4$ MW. 
Figure~\ref{FIG-Beamline} shows the configuration of the beamline. The primary proton beam enters (from the right, in the figure) through a beryllium window sealing off the evacuated beam pipe, and the protons enter the nitrogen gas-filled target chase. After passing through a baffle to protect downstream equipment from a mis-aligned beam, protons impact on a graphite target rod $1.5$-$1.8$ m long, enclosed in a thin-walled titanium alloy vessel. Around 98\% of the protons interact, and high velocity helium gas cooling is employed to remove the $\sim2$\% of beam power that deposits as heat.  The target is embedded in the first horn (shown as A in Fig.~\ref{FIG-Beamline}) which magnetically focuses the secondary particles (pions and kaons) of one selected charge polarity. A second (B) and third (C) horn provide further focusing downstream before the secondaries enter the $194$~m long Decay Pipe, where a large fraction of the pions will decay to neutrinos.  
\subsection{Far Detector Cryostats}
\label{subsec:Cryostats}

\begin{figure}[!tb]
  \centering
  \includegraphics[width=0.8\textwidth]{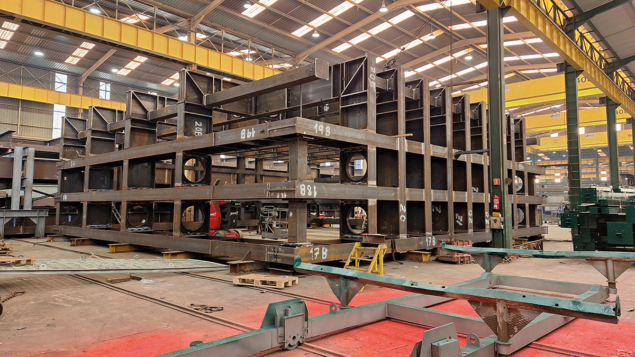}
  \caption{\small 
   The pre-assembly of a section of the second cryostat for the DUNE far detector at the factory in Arteixo, Spain. Credit: CERN.  
  }
  \label{FIG-cryostat}
\end{figure}

\begin{figure}[!tb]
  \centering
  \includegraphics[width=0.9\textwidth]{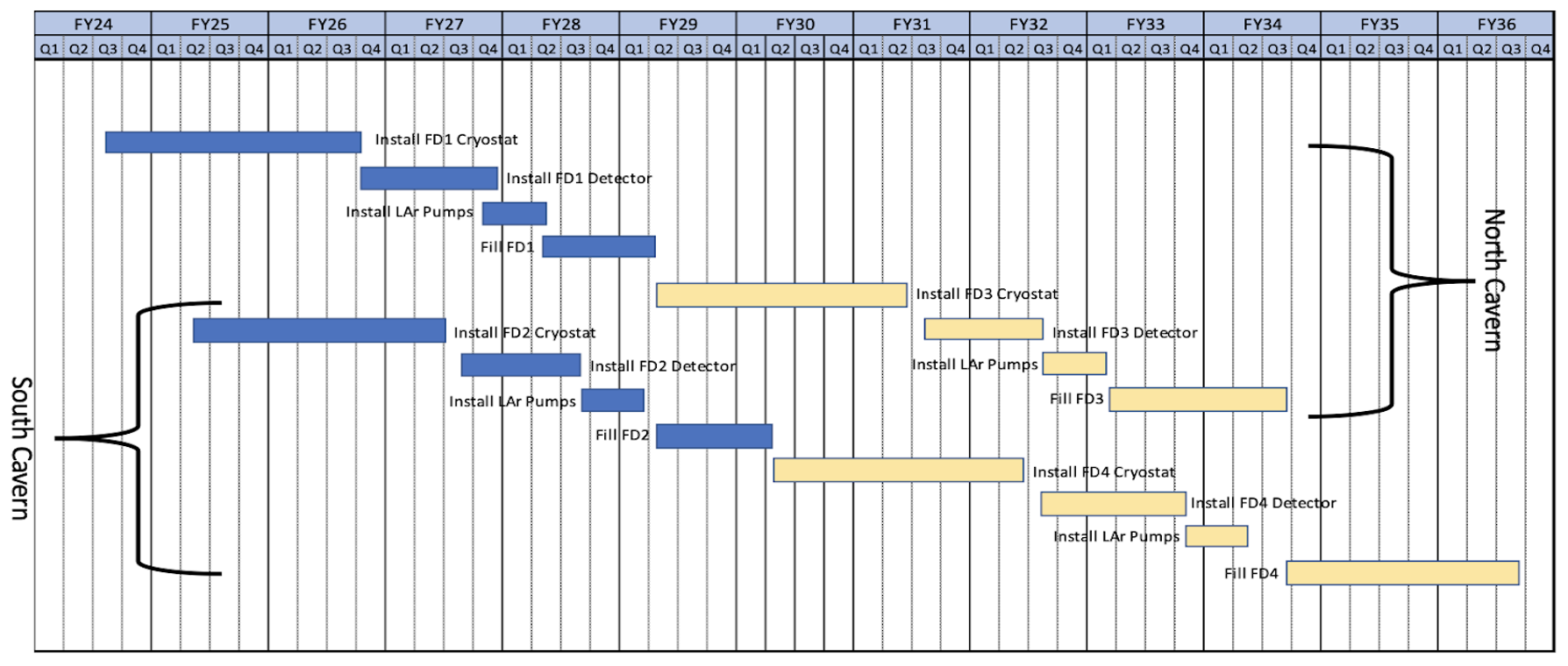}
  \caption{\small 
    A notional, technically limited schedule for DUNE \phasetwo FD modules 3 (FD3) and 4 (FD4) assuming they are vertical drift LAr TPCs similar to DUNE \phaseone FD2. The installation of the FD3 and FD4 cryostats from CERN is expected in 2029 and 2030 respectively.   
  }
  \label{FIG-schedule}
\end{figure}

The DUNE far detector complex will consist of four cryostats, each of which will house one of the four FD modules. Far detector modules 1 and 2 (FD1 \& FD2) and corresponding cryostats are part of the \phaseone construction, and FD modules 3 and 4 (FD3 \& FD4) and corresponding cryostats are part of the \phasetwo construction of the experiment. The cryostats for FD1 and FD2 will be provided by CERN, which will be central to the success of the ongoing execution of DUNE \phaseone. The cost of each cryostat is about \$90 M, resulting in a total contribution of \$180M. The ProtoDUNE program that validates DUNE technologies at the CERN Neutrino Platform has been an extraordinary success, paving the way for CERN to design and engineer cryostats that are 20 times larger for the DUNE FD modules (see Fig.~\ref{FIG-cryostat}). As shown in the schedule in Fig.~\ref{FIG-schedule}, the FD2 and FD1 cryostats will be installed in South Dakota by 2026 and 2027, respectively. 

A defining feature of LBNF/DUNE is the scope of the international partnerships that make this project possible. The significant contributions from European partners to LBNF/DUNE have been critical to the ongoing execution of \phaseone. Given the Collaboration's dedication to realizing the full scope of LBNF/DUNE and the goal of producing sustained world-leading physics, a similar cost-sharing model is anticipated for \phasetwo. Following this, as in \phaseone, the cryostats for FD3 and FD4 are also anticipated to be provided by CERN. As per the technically limited schedule shown in Fig.~\ref{FIG-schedule} for DUNE FD modules, the installation of the FD3 and FD4 cryostats in \phasetwo are expected to start in 2029 and 2030, respectively. As such, successful and timely procurement of the FD3 and FD4 cryostats from CERN will be critical to the success of DUNE \phasetwo.  

\subsection{Neutrino Target}
\label{subsec:Target}
Designing a proton target optimized for neutrino production and which can withstand the shocks experienced from the impact of a multi-MW proton beam, is highly specialized work. The High Power Targets Group and the Project Engineering Group located at STFC's Rutherford Appleton Laboratory (RAL) in the UK are world renowned for their expertise in this area and have a long track record of success in developing targets for T2K, NuMI, NOvA, Mu2e and other experiments. For LBNF/DUNE, the RAL group has taken responsibility to design and deliver a target system capable of running at the initial LBNF beamline power of 1.2 MW plus considerable beamline infrastructure including the beam protection baffle, key components of the target helium cooling circuit and a target exchange system with telemanipulators.    

\begin{figure}[!tb]
  \centering
  \includegraphics[width=0.4\textwidth]{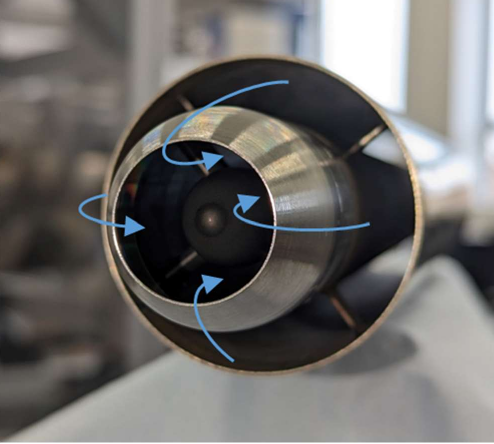}
  \includegraphics[width=0.5\textwidth]{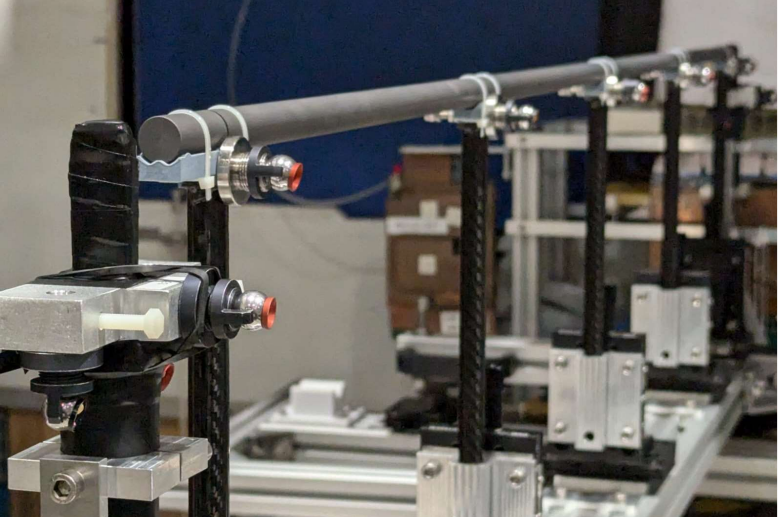}
  \caption{\small 
    (Left) A prototype LBNF target vessel showing the flow lines of He coolant. 
    (Right) A physics replica LBNF graphite target core installed in NA-61 at CERN in 2024.
  }
  \label{FIG-Target}
\end{figure}
The target project has now entered the construction phase and will result in a $1.5$ m long graphite target delivered to FNAL in 2028 ready to be integrated and tested. Some of the most challenging tubular and conical target components have been manufactured (see Fig.~\ref{FIG-Target} (left)) and the technical design of the Target Exchange System is complete and approved for construction.  

The geometry of the target and its support can have a significant effect on the resulting neutrino beam. Specialized simulations of hadron production and transport resulting from the interactions of primary protons with the graphite target, performed by the Warwick University group, have been used to inform the target design choices. The validation of these simulation models has been made possible by running an LBNF replica target in the NA61/SHINE experiment based at CERN. The RAL team installed the replica target at NA61 in June-July 2024 (see Fig.~\ref{FIG-Target} (right)) and collected 1 month of data in July 2024, exposing the target to a proton beam from the CERN SPS at the LBNF nominal energy of $120$ GeV. Such measurements of hadron production at the target are going to ensure that this potential source of systematic uncertainty does not limit the DUNE experiment's measurements of neutrino oscillation parameters. NA61/SHINE will also perform this role for the Hyper-K experiment, and has already proven invaluable to the current generation of oscillation experiments, T2K and NOvA. It is important to the neutrino community that the NA61/SHINE unique facility continues to be supported.     

New studies and development of high power target designs are going to be necessary in order to run reliably with the planned upgraded primary beam power of $2.4$ MW. The Accelerator Complex Evolution (ACE) - Main Injector, Ramp and Targetry (MIRT) initiative from the FNAL Accelerator Group represents a way to reach $>2$ MW beam power by reducing the cycling time in the Main Injector. ACE-MIRT could be ready in 2030, in time for the start-up of beam running at LBNF-DUNE early in the next decade, making this work a priority for the next phase of development led by the RAL Group. Simulation studies will be needed to optimize the target geometry (especially the length) for neutrino production, and a program of materials science is planned to explore options in graphite density for the target and titanium alloys for beam entry and exit windows. These studies will involve a team of specialists working with RAL from across a range of UK institutions: University of Warwick, King's College London, Culham Centre for Fusion Energy, University of Birmingham and the University of Oxford. Funding for this work to continue in the UK is being requested. European support for high power neutrino target development is critical if DUNE is going to be in a position to exploit ACE upgrades for physics.

\section{Summary}
The ability of the DUNE experiment to reach its physics goals is strongly dependent on the provision of the world's highest intensity neutrino beam and a capable facility infrastructure to support the DUNE near and far-detectors.

As host labs, FNAL and SURF have a major role to play, but the delivery of these components would not be possible without a significant investment from the international community and in particular from the European physics community. This submission to the ESPPU has highlighted the European contributions, which range from providing key SRF components for the PIP-II project to high power neutrino targets and far detector cryostats. We emphasize the importance of the delivery of all DUNE \phasetwo elements to achieve full DUNE physics sensitivity and meet P5 goals, particularly reaching a mean sensitivity to leptonic CP violation of better than three standard deviations over more than 75\% of the range of possible values of the CP-violating phase \deltacp. A consequence of this is that there is a strong expectation, and need, for European collaboration to continue and in particular to support the design of a target that can operate beyond $2$~MW beam power and cryostats capable of supporting the DUNE FD3 and FD4 detectors.

\newpage



\bibliographystyle{utphys} 
\bibliography{common/references} 

\end{document}